\documentclass[aps,twocolumn,prl,amsmath,amssymb,showpacs]{revtex4}

\usepackage{graphicx,epsfig}

\begin{document}

\title{
Enhanced thermopower under time-dependent gate voltage}

\author{Adeline Cr\'epieux$^a$}
\author{Fedor $\check{\mathrm{S}}$imkovic$^a$}
\author{Benjamin Cambon$^a$}
\author{Fabienne Michelini$^b$}

\affiliation{$^a$Centre de Physique Th\'eorique, Aix-Marseille Universit\'e, 163 avenue de Luminy, 13288 Marseille, France\\}
\affiliation{$^b$Institut Mat\'eriaux Micro\'electronique Nanosciences de Provence, Aix-Marseille Universit\'e, Avenue Escadrille Normandie Niemen, 13397 Marseille, France}

\date{\today}


\begin{abstract}
We derive formal expressions of time-dependent energy and heat currents through a nanoscopic device using the Keldysh nonequilibrium Green function technique. Numerical results are reported for a metal/dot/metal junction where the dot level energy is abruptly changed by a step-shaped voltage pulse. Analytical linear responses are obtained for the time-dependent thermoelectric coefficients. We show that the Seebeck coefficient can be enhanced in the transient regime up to an amount (here rising $40\%$) controlled by both the dot energy and the height of the voltage step.
\end{abstract}

\maketitle


Since their discoveries in 1821 by Seebeck~\cite{seebeck} and in 1834 by Peltier~\cite{peltier}, thermoelectric effects have been exploited for many applications, such as heat voltage converters, thermocouples or refrigerators. The Seebeck coefficient, or thermopower $S$, measures the voltage induced by a temperature gradient through an open circuit, whereas the Peltier coefficient $\Pi$ measures the heat flow induced by an applied current for no temperature gradient.
In the linear response regime, the Onsager relation gives $\Pi=-S T$, where $T$ is the average temperature of the sample.

The recent achievement of nanoscale systems has invigorated research activities in this field and renewed  the quest of the great thermopower (see Ref.~\onlinecite{dubi1} for a recent review).
On the one hand, stationary Seebeck coefficients have been measured in different nanoscale systems: quantum dot~\cite{scheibner}, atomic-size contacts~\cite{ludoph}, spin valves~\cite{bakker}, nanowires~\cite{hochbaum_boukai}, and carbon nanotube~\cite{suma_small}. The Landauer-B\"uttiker formalism used for the electrical conductance was extended to model thermal transport in microstructures with many terminals~\cite{butcher,sivan}, including inelastic effects~\cite{matveev_galperin}. 
Validity of the conventional thermodynamics linear equations was deeply questioned in mesoscopic systems: the Onsager relations between heat and charge transport coefficients~\cite{butcher,iyoda}, the Wiedemann-Franz law which links electrical and thermal conductances~\cite{vavilov}, and the Fourier law~\cite{dubi2}. 
On the other hand,  time-dependent electric transport also benefits from active research works. Single-electron time-control has been demonstrated experimentally~\cite{feve,lai}, with a fair agreement with earlier theoretical developments~\cite{pastawski,jauho}. More recently, deeper issues as memory effects~\cite{stefanucci} or interplay between multiple time-modulations~\cite{arrachea} have been addressed theoretically. Besides, calculation of time-dependent heat current in a linear phonon chain has recently been achieved~\cite{cuansing}. 
However, thermopower dynamics lacks both experimental as well as theoretical investigations.

\begin{figure}[!h]
       \includegraphics[width=7cm]{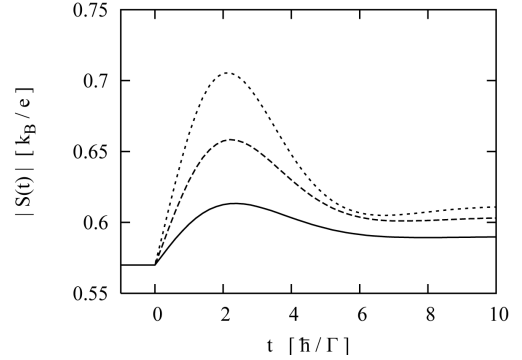}
\caption{Increase of the Seebeck coefficient in the transient regime of a metal/dot/metal junction shown in Fig.~\ref{schema}: $\tilde{\varepsilon}_0=0.5$ with $\tilde{\gamma}_0=0.05$ (solid line), $\tilde{\gamma}_0=0.1$ (dashed line), and $\tilde{\gamma}_0=0.15$ (dotted line). We take symmetric barriers: $\Gamma_R=\Gamma_L$, $t_0=0$, $\varepsilon_F=0$, $k_BT=0.1$. The unit for energies is $\Gamma$.}\label{coefficient}
\end{figure}

\begin{figure}[!h]
       \includegraphics[width=6cm]{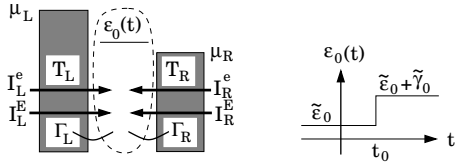}   
\caption{(Left) schematic of metal/dot/metal junction with current directions for each lead; (right) time-dependence of the dot energy level. We define chemical potentials and temperatures of the left and right leads as: $\mu_{L,R}=\varepsilon_F\pm e\Delta V/2$ and $T_{L,R}=T\pm\Delta T/2$. The Fermi energy is $\varepsilon_F$ and $T=(T_R+T_L)/2$ is the average temperature. }\label{schema}
\end{figure}

This letter gives a first insight into time-dependent nonequilibrium thermoelectric transport. As a major result, illustrated in Fig.~\ref{coefficient}, we show that the thermopower can be strongly enhanced during the transient regime in a metal/dot/metal device, presented in Fig.~\ref{schema}. Time evolution of the thermopower exhibits promising features. Indeed, it can be significantly modified and controlled by changing the dot energy from $\tilde{\varepsilon}_0$ to $\tilde{\varepsilon}_0+\tilde{\gamma}_0$ at time $t_0=0$: starting from the stationary value at $t<t_0$, it increases during a finite time, and then it converges toward its new stationary value at $t\rightarrow\infty$.  

Time-dependent heat current through left (L) or right (R) reservoir in equilibrium reads
\begin{eqnarray}\label{thermo}
\langle I_{L,R}^h(t)\rangle= \langle I_{L,R}^E(t)\rangle-\frac{\mu_{L,R}(t)}{e}\langle I_{L,R}^e(t)\rangle~,
\end{eqnarray} 
where $\langle I_{L,R}^E(t)\rangle$ is the energy current, $\langle I_{L,R}^e(t)\rangle$ the electric current, and $\mu_{L,R}$ the chemical potential. 
Time-dependent Seebeck coefficient can be obtained from the ratio between voltage gradient $\Delta V$ and temperature gradient $\Delta T$ between the two reservoirs, when both left and right time-dependent electric currents cancel
\begin{eqnarray}\label{def_seebeck}
S(t)=-\left.\frac{\Delta V}{\Delta T}\right|_{\langle I^e_{L}(t)\rangle=\langle I^e_{R}(t)\rangle=0}~,
\end{eqnarray}
whereas time-dependent Peltier coefficient is defined as
\begin{eqnarray}\label{def_peltier}
\Pi(t)=\left.\frac{\langle I^h_{L}(t)\rangle-\langle I^h_{R}(t)\rangle}{\langle I^e_{L}(t)\rangle-\langle I^e_{R}(t)\rangle}\right|_{\Delta T=0}~.
\end{eqnarray}

The general system we consider consists of $N$ energy levels in an interacting central region connected to non-interacting left and right leads. The total hamiltonian reads $H=H_L+H_R+H_c+H_T$ with
\begin{eqnarray}
H_{L,R}&=&\sum_{k\in L,R}\varepsilon_{k}(t)c^\dag_{k}c_{k}~,\\
H_{c}&=&\sum_n\varepsilon_n(t)d_n^\dag d_n+H_{\mathrm{int}}~,\\
H_T&=&\sum_{p=L,R}\sum_{k\in p,n}V_{kn}(t)c^\dag_{k}d_n+h.c.~,
\end{eqnarray}
where $c^\dag_k$ $(d^\dag_n)$ and $c_k$ $(d_n)$ are the creation and annihilation operators for the leads (dot). $H_{\mathrm{int}}$ is the interacting part of $H_c$ (Coulomb interactions, phonons coupling, etc...). The energy band $\varepsilon_k$ of the reservoirs can be time-dependent through a bias change (source-drain voltage). Energy levels of the central region $\varepsilon_n$ can be time-dependent through a modulation of gate voltage. For the sake of generality, hopping amplitudes $V_{kn}$ are also allowed to be time-dependent. An extension to multiterminal system with additional degrees of freedom, \textit{e.g.} spin, is straightforward. In this calculation, we only consider the electron contribution to the energy current. 

The energy current operator is related to the time derivative of the hamiltonian describing the leads~\cite{wang} by $I^E_{L,R}=-\dot H_{L,R}$. Calculating commutators $[H_{L,R},H]$, we end up with ($\hbar=1$)
\begin{eqnarray}
I^E_{L,R}(t)=&&i\sum_{k\in L,R,n}\varepsilon_{k}(t)V_{kn}(t)c^\dag_{k}d_n+h.c.\nonumber\\
&&-\sum_{k\in L,R}\dot\varepsilon_k(t)c^\dag_{k}c_{k}~.
\end{eqnarray}
Thus, the average energy current reads
\begin{eqnarray}
\langle I^E_{L,R}(t)\rangle=&&2\mathrm{Re}\bigg\{\sum_{k\in L,R,n}\varepsilon_{k}(t)V_{kn}(t)G^<_{nk}(t,t)\bigg\}\nonumber\\
&&-\mathrm{Im}\bigg\{\sum_{k\in L,R}\dot\varepsilon_{k}(t)G^<_{kk}(t,t)\bigg\}~,
\end{eqnarray}
where $G^{<}_{kk'}(t,t')=i\langle c^\dag_{k'}(t')c_k(t)\rangle$ is the lead Green function, and
where we have introduced the mixed Green function $G^<_{nk}(t,t')=i\langle c^\dag_k(t')d_{n}(t)\rangle$ which obeys the Dyson equation
\begin{eqnarray}
G^<_{nk}(t,t')&=&\sum_{n'}\int_{-\infty}^{\infty} dt_1V^*_{kn'}(t_1)
\bigg[G^r_{nn'}(t,t_1)g^<_{k}(t_1,t')\nonumber\\
&+&G^<_{nn'}(t,t_1)g^a_{k}(t_1,t')\bigg]~.
\end{eqnarray}
$G^{<}_{nn'}(t,t')=i\langle d^\dag_{n'}(t')d_n(t)\rangle$ is the dot Green function. $g^<_{k}(t,t')=if(\varepsilon_{k})e^{-i\int_{t'}^t dt_1\varepsilon_{k}(t_1)}$ and $g^a_{k}(t,t')=i\Theta(t'-t)e^{-i\int_{t'}^t dt_1\varepsilon_{k}(t_1)}$ are the Green functions of the isolated leads. Expression of the energy current becomes
\begin{eqnarray}
\label{energy_current}
\langle I^E_{L,R}(t)\rangle=
&&2\mathrm{Re}\bigg\{\mathrm{Tr}\Big\{\int_{-\infty}^{\infty} dt_1\big[{\bf G}_d^r(t,t_1){\bf \Xi}^<_{L,R}(t_1,t)\nonumber\\
&&+{\bf G}_d^<(t,t_1){\bf \Xi}^a_{L,R}(t_1,t)\big]\Big\}\bigg\}\nonumber\\
&&-\mathrm{Im}\bigg\{\mathrm{Tr}\Big\{{\bf \dot E}_{L,R}(t){\bf G}^<_{L,R}(t,t)\Big\}\bigg\}~,
\end{eqnarray}
where we have defined the self-energy associated to energy transfer as
${\bf \Xi}^{a,<}_{L,R}(t,t')=\sum_{k\in L,R}{\bf V}^*_{k}(t)g^{a,<}_{k}(t,t')\varepsilon_{k}(t'){\bf V}_{k}(t')$. In these expressions, ${\bf \dot E}_{L,R}$ and ${\bf V}_k$ are vectors whereas ${\bf G}_d^{r,<}$, ${\bf G}_{L,R}^{<}$, ${\bf \Xi}^{a,<}_{L,R}$ are matrices. Last term in Eq.~(\ref{energy_current}) is a pure reservoir contribution to the energy current.

We now consider a non-interacting metal/dot/metal junction with a single dot level $\varepsilon_0$ connected to reservoirs with constant energy bands. This model is suitable for experiments in which Coulomb interaction is weak in the dot with strong coupling to reservoirs \cite{feve,gabelli}. In that case, the energy current takes the form
\begin{eqnarray}\label{th_current}
&&\langle I^E_{L,R}(t)\rangle=2\mathrm{Re}\bigg\{\int_{-\infty}^{\infty} dt_1 \int_{-\infty}^{\infty} \frac{d\varepsilon}{2\pi}i\varepsilon e^{i\varepsilon(t-t_1)}\nonumber\\
&&\times\Gamma_{L,R}(\varepsilon,t_1,t)\big[G_d^r(t,t_1)f_{L,R}(\varepsilon)+G_d^<(t,t_1)\Theta(t-t_1)\big]\bigg\}~,\nonumber\\
\end{eqnarray}
where $\Theta$ is the Heaviside function, $f_{L,R}$ the Fermi-Dirac distribution function, $\rho_{L,R}$ the density of states, and $\Gamma_{L,R}(\varepsilon_k,t,t')=2\pi\rho_{L,R}(\varepsilon_k)V^*_{k}(t)V_{k}(t')$ 
measures the strength of the coupling between the dot and each lead.

We investigate the time-dependent thermoelectric response to a unique change $\varepsilon_0(t)=\tilde\varepsilon_0+\gamma_0(t)$ with $\gamma_0(t)=\tilde{\gamma}_0\Theta(t-t_0)$. This models a dot energy switching from $\tilde{\varepsilon}_0$ to $\tilde{\varepsilon}_0+\tilde{\gamma}_0$ by applying a gate voltage at $t_0$ (see Fig.~\ref{schema}). The time-dependent heat current defined by Eq.~(\ref{thermo}) for $p=L,R$ is now expressed in terms of the spectral function $A(\varepsilon,t)$ as
\begin{eqnarray}\label{time_heat_current}
&&\langle I^h_p(t)\rangle=-\frac{1}{h}\Gamma_{p}\bigg[2\int_{-\infty}^{\infty}(\varepsilon-\mu_p) f_{p}(\varepsilon)\mathrm{Im}\{A(\varepsilon,t)\} d\varepsilon\nonumber\\
&&+\sum_{p'=L,R}\Gamma_{p'}\int_{-\infty}^{\infty} (\varepsilon-\mu_{p'})f_{p'}(\varepsilon)|A(\varepsilon,t)|^2 d\varepsilon
\bigg]~,
\end{eqnarray}
with~\cite{jauho}
\begin{eqnarray}\label{spectral_function}
A(\varepsilon,t)=\frac{\varepsilon-\tilde{\varepsilon}_0+i\Gamma/2-\tilde{\gamma}_0e^{i(t-t_0)(\varepsilon-\tilde{\varepsilon}_0-\tilde{\gamma}_0+i\Gamma/2)}}{(\varepsilon-\tilde{\varepsilon}_0+i\Gamma/2)(\varepsilon-\tilde{\varepsilon}_0-\tilde{\gamma}_0+i\Gamma/2)}~,
\end{eqnarray}
and  $\Gamma=\Gamma_{L}+\Gamma_{R}$, for which we assume that $\Gamma_{p}$ does not depend on energy, and simply reduces to $\Gamma_{p}=2\pi\rho_{p}|V_{p}|^2$.

\begin{figure}[!h]
       \includegraphics[width=6cm]{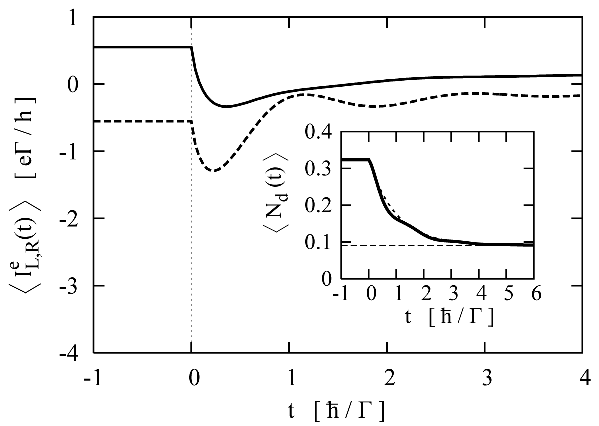}
       \includegraphics[width=6cm]{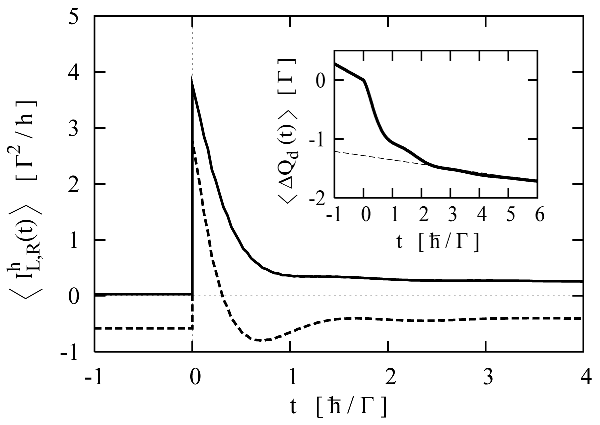}   
\caption{(Top graph) electric current and (bottom graph) heat current through left lead (solid lines) and right lead (dashed lines) as a function of time $t$, for $\Gamma_R=\Gamma_L$, $t_0=0$, $\tilde{\varepsilon}_0=0.5$, $\tilde{\gamma}_0=2.5$, $k_BT_L=1$, $k_BT_R=0$, $\mu_{L,R}=\pm 0.5$. Insets show (top inset) the dot occupation, and (bottom inset) the dot heat. Dashed lines indicate the stationary limits at $t\rightarrow \infty$.}\label{time_dependent}
\end{figure}

The integration over energy in Eq.~(\ref{time_heat_current}) has been performed numerically. In Fig.~\ref{time_dependent} is plotted the time evolution of electric and heat currents when the dot energy is modified abruptly at $t_0=0$. Starting from constant values at $t<0$, left and right currents converge toward constant values at $t\rightarrow \infty$. Between these two limits, currents show strong time-dependent variations. Currents through the zero-temperature, $T_R=0$, right lead (, see dashed lines in Fig.~\ref{time_dependent}) exhibit time oscillations whose period is related to $\tilde{\varepsilon}_0$ and $\tilde{\gamma}_0$, as Eq.~(\ref{spectral_function}) explicitly indicates. These oscillations of the electric current have been measured through a Ge dot~\cite{lai}. Concerning the heat current, experimental results are still needed. In the left lead (see solid lines in Fig.~\ref{time_dependent}), these oscillations disappear due to thermal effects given by  $T_L\neq0$. 
 
Using particle number conservation, the average dot occupation number is calculated from electric currents as $\langle N_d(t)\rangle=e^{-1}\int \langle I_{dis}^e(t)\rangle dt$, where $ I_{dis}^e(t)=I_L^e(t)+I_R^e(t)$ is the displacement current \cite{lai}. 
In the top inset of Fig.~\ref{time_dependent}, $\langle N_d(t)\rangle$ globally follows an exponential decrease $\langle N_d(\infty)\rangle(1-e^{-t/\tau_r})+\langle N_d(0)\rangle e^{-t/\tau_r}$ (see dotted line) characterized by the relaxation time $\tau_r=\hbar/\Gamma$: weaker is the coupling between the dot and the leads, longer is the relaxation time. Time evolution of $\langle N_d(t)\rangle$ shows oscillations around this decrease, that have been already observed in experiments \cite{lai}.

Eq.~(\ref{thermo}) comes from thermodynamic relations in leads at equilibrium: $dH_{L,R}=dQ_{L,R}+\mu_{L,R} dN_{L,R}$, where $N_{L,R}$ is the lead occupation number and $Q_{L,R}$ is the lead heat. Similarly, for the dot  out-of-equilibrium, we write $dH_c=dQ_d-\mu_{L} dN_{L}-\mu_{R} dN_{R}$: the energy change in the dot  reflects a balance between heat variation and charges leaving the dot times their energies. Thus, we define and numerically calculate an average heat in the dot as $\langle Q_d(t)\rangle=\int \langle I_d^h(t)\rangle dt$, where $I_d^h(t)=I_L^h(t)+I_R^h(t)-\dot H_T(t)$. This definition perfectly agrees with energy conservation including contribution from tunneling. 
In the bottom inset of Fig.~\ref{time_dependent}, the time evolution of $\Delta \langle Q_d(t)\rangle=\langle Q_d(t)\rangle-\langle Q_d(0)\rangle$ shows a behavior similar to the dot occupation number  (top inset of Fig.~\ref{time_dependent}). However, dramatic differences occur in the stationary regimes, for which $\langle N_d(t)\rangle$  is always constant: the dot heat varies linearly in time, without any violation of conservation laws.

In the linear response limit, the time-dependent Seebeck coefficient, defined by Eq.~(\ref{def_seebeck}), can be obtained from the approximate Fermi-Dirac distribution function~\cite{butcher}: $f_{L,R}(\varepsilon)\approx f_0(\varepsilon)+f_0'(\varepsilon)(\mu_{L,R}-(\varepsilon-\varepsilon_F)T_{L,R}/T))$, where $f_0$ is the Fermi-Dirac distribution function for the leads when $\mu_L=\mu_R$. Taking left and right electric currents equal zero, we obtain the linear response for $S(t)$ in the case of strong coupling to reservoir and small energy variation:
\begin{eqnarray}\label{time_seebeck}
S(t)=-\frac{\int_{-\infty}^{\infty}d\varepsilon f_0'(\varepsilon)(\varepsilon-\varepsilon_F){\mathcal T}(\varepsilon,t)}{eT\int_{-\infty}^{\infty} d\varepsilon f_0'(\varepsilon){\mathcal T}(\varepsilon,t)}~,
\end{eqnarray}
where ${\mathcal T}(\varepsilon,t)=-4\Gamma_L\Gamma_R\mathrm{Im}\{A(\varepsilon,t)\}/\Gamma$ is the time-dependent transmission coefficient. 
This result is a generalization of the Seebeck coefficient expression obtained in the stationary case~\cite{butcher,dubi1} including time-dependence of the transmission coefficient. For the steady-state, we have $\langle I^e_L\rangle=-\langle I^e_R\rangle$ (constant $\langle N_d(t)\rangle$), as can be seen in the top graph of  Fig.~\ref{time_dependent}. But in the time-dependent case, $\langle I^e_L(t)\rangle=0$ does not imply $\langle I^e_R(t)\rangle=0$ because of the displacement current. Since the Seebeck coefficient is measured in an open circuit, we must find the adequate $\Delta V$ and $\Delta T$ which simultaneously cancel both currents. It is important to emphasize that Eq.~(\ref{time_seebeck}) is only valid under the following assumptions: i) linear response (small $\Delta V$ and $\Delta T$ in comparison to $\tilde{\varepsilon}_0$ and $T$); ii) high transmission through the barriers (large $\Gamma_L$ and $\Gamma_R$ in comparison to other energies); and iii) small gate voltage time-variation $\tilde{\gamma}_0$, in comparison to $\tilde{\varepsilon}_0$. 
Indeed, these assumptions allow to approximate $|A(\varepsilon,t)|^2\approx-2\mathrm{Im}[A(\varepsilon,t)]/\Gamma$, and, hence, to cancel both $\langle I^e_L(t)\rangle$ and $\langle I^e_R(t)\rangle$ at any time. 

Similar expressions can be obtained for the time-dependent Peltier coefficient defined by Eq.~(\ref{def_peltier})
\begin{eqnarray}\label{time_peltier}
\Pi(t)=\left.\frac{\int_{-\infty}^{\infty}d\varepsilon (\varepsilon-\varepsilon_F)(f_L(\varepsilon)-f_R(\varepsilon)){\mathcal T}(\varepsilon,t)}{e\int_{-\infty}^{\infty} d\varepsilon (f_L(\varepsilon)-f_R(\varepsilon)){\mathcal T}(\varepsilon,t)}\right|_{\Delta T=0}~,\nonumber\\
\end{eqnarray}
 following the same assumptions.
In the linear response regime, Eqs.~(\ref{time_seebeck}) and (\ref{time_peltier}) verify the Onsager relation $\Pi(t)=-TS(t)$ at any time. 

\begin{figure}[!h]
       \includegraphics[width=5.5cm]{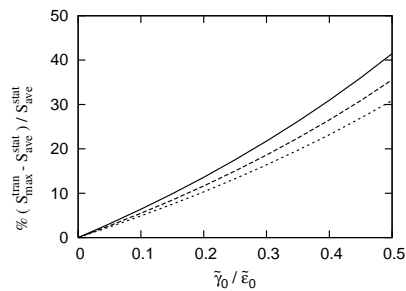}
\caption{Percentage of increase of the Seebeck coefficient maximum in the transient regime $S^{\mathrm{tran}}_{\mathrm{max}}$ as a function of $\tilde{\gamma}_0/\tilde{\varepsilon}_0$ for $k_BT=0.05$ (solid line), $k_BT=0.15$ (dashed line) and $k_BT=0.5$ (dotted line). We take $\Gamma_R=\Gamma_L$ and $\varepsilon_F=0$.}\label{seebeck_max}
\end{figure}

Fig.~\ref{coefficient} is obtained using Eq.~(\ref{time_seebeck}) of the linear response. It is shown an increase of the thermopower after a step-shaped gate-voltage pulse was applied. The reason is that in the transient regime, the system is much more sensitive to  temperature or electrostatic variations. 
Furthermore, we measure the thermoelectric benefit of the transient regime calculating the percentage 
 $(S^{\mathrm{tran}}_{\mathrm{max}}-S^{\mathrm{stat}}_{\mathrm{ave}})/S^{\mathrm{stat}}_{\mathrm{ave}}$ where $S^{\mathrm{tran}}_{\mathrm{max}}$ is the maximum value of $S$ and $S^{\mathrm{stat}}_{\mathrm{ave}}=(S(t<t_0)+S(t\rightarrow\infty))/2$.
In Fig.~\ref{seebeck_max}, we have plotted the percentage of thermopower increase as a function of $\tilde{\gamma}_0/\tilde{\varepsilon}_0$. This ratio plays an important role since it does control the thermopower increase. In such a junction, the transient thermopower can be tuned by both $\tilde{\varepsilon}_0$ and $\tilde{\gamma}_0$, which depend on the dot structural properties and of the applied gate voltage (see Fig.~\ref{schema}).
Higher is the ratio, higher is the thermopower increase. Here, an increase up to $40\%$ is obtained at small temperature.

We have proposed a first approach to heat dynamics in nanoscale junctions. General formula for the time-dependent heat and energy currents flowing through an interacting resonant-tunneling system have been obtained. We show that an enhanced thermopower can be generated during the transient regime in a metal/dot/metal junction, and that its maximum value can be tuned by both the dot energy and the gate voltage.  With such numerical investigations, it will be possible to go beyond the linear response for the Seebeck and Peltier coefficients, and further determine non-linear thermodynamic laws. Moreover, we shall consider interacting systems in order to analyze phonon bath contribution \cite{cuansing}, impact of  electron-phonon interaction \cite{matveev_galperin}, and to study influence of charging effects in other experiments \cite{splettstoesser}.

A.C. thanks I. Safi for discussion and E. Bernardo for her help in bibliographic researches.


\end{document}